\documentclass[10pt]{iopart}

\usepackage[english]{babel}
\usepackage{graphicx}
\usepackage{iopams,cite}

\newcommand{\pderv}[2]{\frac{\partial #1}{\partial #2}}
\newcommand{\pdervi}[2]{\partial #1\slash\partial #2}

\newcommand{\rem}[1]{}

\newcommand{\ind}[1]{{\mathrm{#1}}}

\newcommand{\new}[1]{}

\def\rmd{\mathrm{d}}

\def\oc{\omega_\ind{c}}

\def\ob{\omega_\ind{b}}
\def\o{\omega}
\def\I{I_\perp}
\def\E{\mathcal{E}}
\def\K{\mathcal{K}}
\def\res{\ind{r}}

\begin{document}
\title
{Towards explanation of the two-frequency heating effect in electron cyclotron ion sources}
\author{A G Shalashov, E D Gospodchikov, I V Izotov, V~A~Skalyga} 
\address{Institute of Applied Physics of Russian Academy of Sciences, 46 Ulyanov Str., 603950, Nizhny Novgorod, Russia}
\ead{ags@ipfran.ru}

\date{\today}

\begin{abstract}
The quasilinear model of electron cyclotron resonance (ECR) heating in a mirror magnetic field predicts essential broadening of the electron distribution function in case of bichromatic wave. This may stabilize kinetic instabilities in the electron-cyclotron frequency range resulting in improved performance of ECR ion sources, the effect which is observed in many independent experiments. 
\end{abstract}



\ioptwocol
\maketitle

\section{Introduction}

Electron cyclotron resonance (ECR) ion sources play essential role in fundamental nuclear physics research and applications, what motivates further development and optimization of such devices, see \cite{ECRrev} and references therein. One of the essential factors limiting the performance of modern ECR ion sources, in particular the extracted high charge state beam currents, is attributed to development of kinetic plasma instabilities in the ECR frequency range --- particle ejections, which are inherent to the burst regime of the cyclotron instability, cause oscillations of the plasma potential and the beam current accompanied with a significant decrease of the average ion charge \cite{izo1,izo2,izo3,olli_2015}. 
The control of high-frequency instabilities in ECR ion sources is possible either careful tuning of the plasma heating conditions by a monochromatic wave \cite{b-prl, b-epl, b-prap} or by adding the second wave at offset frequency \cite{izo3}. 
The latter solution seems to be very natural since the most experiments aimed at development of high-performance ECR ion sources have an option of two or more frequency heating as a consequence of gradual upgrade of the heating systems when old sources remain operating together with new ones, at different frequency. In all cases two-frequency heating leads to increase of the source performance, especially the high charge state ion productin \cite{tf1,tf2,tf3,tf4,tf5}; in some cases a setup operates normally without a second heating frequency only in an extremely narrow  range of parameters \cite{susi}. 

This very reproducible effect has been found occasionally and still lacks generally accepted  theoretical explanation. In the present paper we support the idea, originally proposed in  \cite{izo3}, that two-frequency heating results in stabilization or essential reduce of the electron-cyclotron instability and thus postpone its negative consequences. As discussed hereafter, the quasilinear model of electron cyclotron plasma-wave interaction provides arguments that two-frequency heating would generate electrons with naturally more stable distribution function.

\section{Quasilinear diffusion equation}

In this paper we follow the quasi-linear theory adopted for plasma confined in a laboratory mirror trap as formulated in \cite{bible,bible2}; so we ask a reader to refer this review work for the related references on original (historical) works resulted in such formalism. 
In an open magnetic trap, electrons are involved in two oscillatory motion: a cyclotron gyration around the magnetic field and bouncing along the magnetic field line between the mirrors; drifts across the magnetic field may be neglected for the fast time-scales considered here.  Thus, we assume a standard ordering of the drift theory, 
\begin{equation*}\tau_\ind{c}\ll \tau_\ind{b}\ll\tau_\ind{dr},\end{equation*}
where 
\begin{equation*}\tau_\ind{c}=2\pi/\oc\;, \tau_\ind{b}=2\pi/\ob\sim (l/\rho_\ind{c})\,\tau_\ind{c},\;\tau_\ind{dr}\sim(l/\rho_\ind{c})^2\, \tau_\ind{c}\end{equation*} 
are, correspondingly,  characteristic times of the gyromotion, bounce oscillations and drift across the magnetic field, $\rho_\ind{c}$ is the electron Larmor radius and $l$ is inhomogeneity scale of the magnetic field.  
Omitting the transverse drifts, one may consider two degrees of freedom of a single electron---the phase of cyclotron gyration $\varphi$ and the coordinate $z$ along the field line; $z=0$ stands for the minimum of the magnetic field strength. Canonically conjugated momenta are, correspondingly, the transverse adiabatic invariant $\I$ and the longitudinal momentum $p_z$. Here ``transverse'' and ``longitudinal'' are related to the direction of the external magnetic field, 
\begin{equation*}\I={p_\perp^2}/{2m\oc(z)},\qquad\oc(z)={eB(z)}/{mc}, \end{equation*}
$\oc(z)$ is the non-relativistic electron-cyclotron frequency which depends on $z$,  $m$ is the electron rest mass, and $e$ is the elementary charge. In the absence of electromagnetic waves, an electron motion is fully characterized by conservation of the adiabatic invariant $\I$ and the electron energy $\E$, 
\begin{eqnarray*}\fl
\E\equiv mc^2(\gamma-1)=\nonumber \\ =\sqrt{m^2c^4+c^2p_z^2 +2mc^2 \oc(z)\vphantom{^1}\,\I}-mc^2.\label{eqE}
\end{eqnarray*}
Non-perturbed electron distribution function is an  arbitrary function of two invariants of motion, $f(\E,\I)$, normalized over a total number of particles in a magnetic flux tube as
\begin{equation}\label{eqN}\int f \,\rmd\Gamma=N, \qquad\rmd\Gamma=\rmd\E\rmd\I. \end{equation}
In this representation, the distribution function does not depend on $z$ and conserves during the bounce motion between the magnetic mirrors.

Now let us assume first that the  bouncing electron interacts with a \emph{monochromatic }electromagnetic field  with frequency $\o$. The interaction takes place in the vicinity of some point $z_\res$ along the magnetic field line where the electron-cyclotron resonance condition is met. Using a quantum mechanic analogy, we may state that the interaction with one photon would result in simultaneous change of the electron energy and transverse momentum as $\Delta\E=\!\pm\hbar\o$ and $\Delta\I=\pm s\hbar$, where different signs correspond to either absorption or emission of a photon and  $s$ is the integer number of cyclotron harmonic. Correspondingly, variation of the electron distribution function during one elementary act is
\begin{equation*}\Delta (f \rmd\Gamma)=\left(\Delta\E\pderv{}{\E}+\Delta\I\pderv{}{\I}\right)(f\rmd\Gamma). \end{equation*}

Having in mind an application to ECR ion source, we may simplify further notations assuming the fundamental harmonic $s=1$ hereafter. 
Combining  $\Delta(\E-\o\I)=0$, one finds that  
\begin{equation}\label{eqK}\E-\o\I=\mathrm{const}\end{equation} 
is conserved during the wave-particle interaction.
The resonant electrons exhibit a classic one-dimensional random walk motion along the curve \eref{eqK} in $(\E,\I)$-space. A slow evolution of the electron distribution function during many interaction acts may be described by the Fokker--Planck equation 
\begin{equation}\label{eqQ1}\pderv{f}{t}=\left(\pderv{}{\E}+\frac1\o\pderv{}{\I}\right)\frac{ D_\ind{ql}}{\ob}\left(\pderv{}{\E}+\frac1\o\pderv{}{\I}\right) (\ob f), \end{equation}
where $\ob(\E,\I)$ is the frequency of bounce-oscillations, an additional multiplier due to variation of the elementary volume $\rmd\Gamma$ in our non-canonical phase-space that have been erroneously omitted in \cite{bible} but taken into account in \cite{bible2}, 
$D_\ind{ql}(\E,\I)$ is the quasilinear diffusion coefficient, and \eref{eqK} is a quasilinear diffusion curve. 
With new variables 
\begin{equation*}
\xi=(\E+\o\I)/2,\qquad \K=(\E-\o\I)/2,
\end{equation*}
equation \eref{eqQ1} transforms to 
\begin{equation}\label{eqQ2}\pderv{\tilde f}{t}=\pderv{}{\xi}\left(\frac{ D_\ind{ql}}{\ob}\pderv{(\ob\tilde f)}{\xi}\right), \end{equation}
a true one-dimensional form in which the quasilinear diffusion occurs over the axis $\xi$ while  $\K$ is a conserving label of the diffusion curve. Here we consider distribution function $\tilde f(\xi,\K)\equiv f(\E,\I)$ with the  elementary volume 
$\rmd\Gamma=\rmd\xi\rmd\K/\o$ (same norm  as \eref{eqN}). 
Following general approach to Fokker--Planck process, the quasilinear diffusion may be defined as
$D_\ind{ql}=\left\langle (\Delta\xi)^2\right\rangle/2\tau_\ind{b}$, 
where $\tau_\ind{b}=2\pi/\ob$ is a mean time between consecutive ``kicks'' with dispersion $\langle (\Delta\xi)^2\rangle$, and $\langle...\rangle$ denotes the  bounce-averaging over non-perturbed electron trajectory, i.e. 
\begin{equation*}\left\langle \dots\right\rangle=\frac\ob{2\pi}\oint\dots\frac{\rmd z}{v_z},\quad 
{\tau_\ind{b}}=\oint\frac{\rmd z}{v_z},\quad v_z=\pderv{\E}{p_z}.\end{equation*}

To describe bichromatic heating by two waves with frequencies $\o$ and $\o'$,  we introduce two independent sets of  variables linked to each frequency as 
\begin{eqnarray*}
\xi=(\E+\o\I)/2,\qquad &\xi'=(\E+\o'\I)/2,\, \\
\K=(\E-\o\I)/2,\qquad &\K'=(\E-\o'\I)/2,\,
\end{eqnarray*}
and consider the Fokker-Plank equation with two additive quasilinear operators:
\begin{equation}\label{eqDQ1}\pderv{\tilde f}{t}=\underbrace{\pderv{}{\xi}\left(\frac{ D_\ind{ql}}{\ob}\pderv{(\ob \tilde f)}{\xi}\right)}_{\K=\mathrm{const}}+\underbrace{\pderv{}{\xi'}\left(\frac{ D'_\ind{ql}}{\ob}\pderv{(\ob \tilde f)}{\xi'}\right)}_{\K'=\mathrm{const}}. \end{equation}
Each of these sets alone describes one-dimensional diffusion along conserving $\K$ and $\K'$, however acting together they result in a two-dimensional diffusion hereafter considered in more detail. Note that \eref{eqDQ1} is obtained in a phenomenological way, simply by adding two independent processes. However, this is in good consistence with the general quasilinear approach which implies that even for a single frequency there is no correlation between the consecutive ``kicks'', so adding a second frequency just adds new ``kicks''. Basing on a general experience one may suggest that spectrum broadening would always ease applicability of the quasilinear approach for almost all resonant particles.  Of course, exact definition of that applicability conditions require consideration of the problem from first principles.

\section{Monochromatic heating}\label{secmh}

In a high-performance ECR ion source, the main power is primary deposited into fast electrons. Such electrons, accelerated due to interaction with externally generated waves, originate from a relatively cold and dense background plasma. The energy gained by a typical electron is much larger than its initial energy: from tens keV to MeVs compared to the ionization energy of few tens eV \cite{izot-new}. In the first approximation, one can assume zero initial energy of accelerated electrons. 
Then, for monochromatic heating at  frequency $\o$,  the quasilinear model predicts that resonant electrons are accelerated along the one-dimensional diffusion line $\K=0$ based on zero energy, i.e.
\begin{equation}\label{K0}\E=\o\,\I.
\end{equation} 
Later we will relax this condition allowing some distribution of seed electron energies. 

\begin{figure}[tb]
\centering \includegraphics[width=0.35 \textwidth]{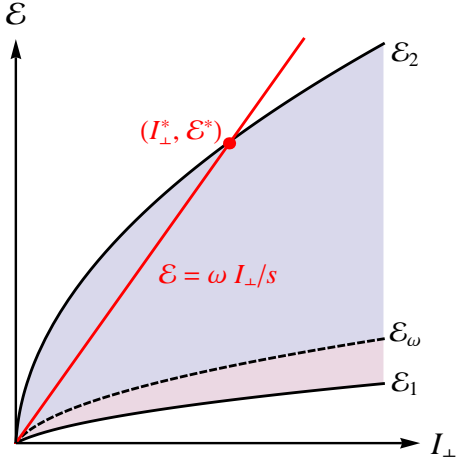}
\caption{Domain of confined particles in a plane formed by the invariants of bounce-motion. Red line shows the quasilinear diffusion curve for particles with zero initial energy.} \label{fig-2}
\end{figure}

On the other hand, the electron energy varies in the range
\begin{equation}\label{Econ}\E_1\le\E\le\E_2,\;\;\E_{1,2}=\sqrt{m^2c^4+2mc^2 \o_{1,2}\vphantom{^1}\,\I}-mc^2,
\end{equation} 
where $\o_1=\min\oc(z)$ and $\o_2=\max\oc(z)$ along a field line. This is illustrated in figure \ref{fig-2}. The lower boundary corresponds to the particles that have only transverse (to the magnetic field) velocity in the trap center (minimum of $B$). The upper boundary denotes the loss-cone---it corresponds to the particles that have zero parallel velocity at the magnetic mirror (maximum of $B$); the particles with $\E>\E_2$ are not reflected by the mirrors. Here we assume that not-confined electrons are not involved in the heating that needs many bounce-oscillations. 
For non-relativistic energies these boundaries become $\E_{1,2}\approx\o_{1,2}\I$. Comparing with \eref{K0}, we see that the effective ECR heating at the fundamental harmonic is possible only if 
\begin{equation}\label{wcon}\o_1\le\o\le\o_2,\end{equation} 
i.e.\ if the cold cyclotron resonance condition is met somewhere along the field line. Once this condition is met, the acceleration goes until $\o\,\I\le\E_2$. In other words, the acceleration stops when the quasilinear diffusion pushes an electron into the loss-cone. This occurs at the point $(\E^*,\I^*)$,
\begin{equation*}\E^*=2mc^2(\o_2/\o-1),\;\; \I^*=\E^*/\o.
\end{equation*} 
Note that this is an essentially relativistic effect related to the mass dependence on momentum.

\begin{figure}[tb]
\centering \includegraphics[width=0.35 \textwidth]{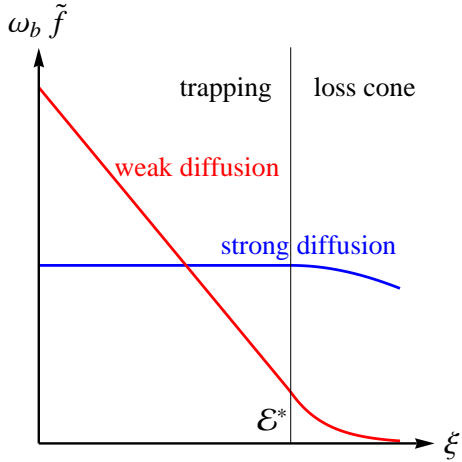}
\caption{Schematic view of the one-dimensional distribution function in the weak and strong regimes of quasilinear diffusion.} \label{fig-2a}
\end{figure}

To characterize the distribution function of fast electrons, let us return to the one-dimensional quasilinear equation \eref{eqQ2}. We want to consider a stationary solution of this equation, so it must be upgraded by adding the particle source  and losses as
\def\nuf{\nu_\ind{lc}}
\def\tuf{\tau_\ind{lc}}
\begin{equation}\label{eqQ3}\pderv{}{\xi}\left(\frac{D_\ind{ql}}{\ob}\pderv{(\ob\tilde f)}{\xi}\right)+S-L=0. 
\end{equation}
Here  $S$ is a source of seed (low-energy) electrons,
\begin{equation*} S= {S_0}\,\delta(\xi)\,\delta(\K),\;\;\int S\,\rmd\Gamma=\frac{S_0}{\o},
\end{equation*}
and losses $L$ are defined as a particle flux through the loss-cone,
\begin{equation*}L= \cases{0& if $0<\xi<\E^*$\\ \tilde f/\tuf \quad & if $\xi\ge\E^*$},\end{equation*} 
with ${\tuf}$ being a characteristic life-time of a electron inside the loss-cone. 
Note that this equation is singular at the loss-cone boundary as  $\ob,D_\ind{ql}\to0$ and $\tuf\to\infty$, however their combinations $D_\ind{ql}/\ob$ and $ \tuf\ob$ remain finite. Therefore, one may consider \eref{eqQ3} as an equation for $F=\ob\tilde f$ that is regular at the loss-cone boundary. Function $F$ is indeed a distribution function in the canonical action-angle variables \cite{bible2},

Now one may consider trapped ($0<\xi\le\E^*$) and lost ($\xi>\E^*$) particles independently assuming continuous distribution function $F$ and its first derivative at the loss-cone boundary $\xi=\E^*$. Having in mind zero losses for the trapped particles, we find a formal solution of \eref{eqQ3} for $0<\xi\le\E^*$ as
\begin{eqnarray}\fl
D_\ind{ql}\,\pderv{(\ob\tilde f)}{\xi}+S_0\ob\delta(\K)=0\;\Rightarrow \nonumber \\ \qquad \ob\tilde f=\left(f^*-{S_0}\int_{\E^*}^{\xi}\frac{\ob(\xi')\rmd\xi'}{D_\ind{ql}(\xi')}\right)\delta(\K),\label{eqQ4}
\end{eqnarray}
where $f^*$  is yet unknown value of $\ob\tilde f$ at the loss-cone boundary $\xi=\E^*$. 
To find analogous solution inside the loss-cone, we neglect the variation of $D_\ind{ql}/\ob$ and $\ob \tuf$ --- this is a valid assumption because the distribution function decreases rapidly with the energy, $\tilde f\to0$ as $\xi\to\infty$. Then for $\xi\ge\E^*$:
\begin{eqnarray}\fl
\pderv{\vphantom{\partial}^2(\ob\tilde f)}{\xi^2}-(\ob\tilde f)/\Delta\xi^2=0\;\Rightarrow \nonumber \\ \qquad
\ob\tilde f=f^* \exp\Big(-(\xi-\E^*)/\Delta\xi\Big)\,\delta(\K),\label{eqQ4a}
\end{eqnarray}
where   $\Delta\xi=\lim_{\xi\to\E^*}\sqrt{D_\ind{ql}\tuf}$ has a clear physical interpretation as a diffusion path inside the loss region. 
Matching these two solutions at $\xi=\E^*$ one finds that 
\begin{equation*}f^*=S_0 \lim_{\xi\to\E^*}\sqrt{\frac {\tuf\ob }{D_\ind{ql}/\ob}}\approx
\frac{S_0\ob^*}{\E^*}\sqrt{\tuf/ \tau_\ind{ql}},\end{equation*} so our stationary solution is fully defined.
To get the last approximate equality we additionally assume $D_\ind{ql}\approx(\E^*)^2/\tau_\ind{ql}$ where $\tau_\ind{ql}$ is a characteristic time of quasilinear diffusion.

Alternatively, one may exclude the loss-cone region from consideration by setting a proper boundary condition at  $\xi=\E^*$. From \eref{eqQ4a} it follows that 
\begin{equation}\label{eqQ4b}
\E^*\pderv{(\ob\tilde f)}{\xi}+\sqrt{\tau_\ind{ql}/\tuf}\,({\ob\tilde f})=0  
\quad[\mathrm{at}\;\xi=\E^*].\end{equation}
As a reasonable approximation, one can use this condition even for non-stationary problems \cite{bible}. 
This condition allows us to distinguish two important limiting cases, both faced when describing a laboratory experiment, illustrated qualitatively in \fref{fig-2a}.

First case is the  regime of ``weak diffusion'' characterized with $\tuf\ll{\tau_\ind{ql}}$. Physically this means that electrons leave the loss-cone faster than they are delivered there under action of waves. In this case the distribution function \eref{eqQ4} is of ``triangular'' shape with  $f^* \ll\mathrm{average}\,(\ob\tilde f)$. A truly empty loss-cone is also a reasonable assumption---when not thinking on conservation of particles one can simply put $f^*=0$, this corresponds to the boundary condition \eref{eqQ4b} with $\tuf\to0$.

In the opposite case of ``strong diffusion'', ${\tau_\ind{ql}}\ll{\tuf}$, electrons are supplied into the loss region faster than they could escape it. This results in effective mixing of particles along the whole diffusion line with approximate conservation of its number. This process is frequently called as a formation of ``the quasilinear plateau'' on the distribution function. The  boundary condition \eref{eqQ4b} then corresponds to (approximately) zero particle flux into the loss-cone, $\pdervi{(\ob\tilde f)}{\xi}=0$ for $\tau_\ind{ql}\to0$. Formally this means that the main contribution to the distribution function  \eref{eqQ4} comes from the dominating constant first term: $\ob\tilde f\approx f^*\delta(\K)$.
Using the normalizing condition \eref{eqN} one can link the plateau height with the total number of particles
\begin{equation*}  {N}=f^*\int_0 ^{\xi^*} \frac{\rmd \xi'}{\o\ob(\xi',0) }. \end{equation*}
Further, using conditions $\K=0$ and $\xi\ll\xi^*$, one may find that $\ob\propto \sqrt{\xi}$ independent of a particular profile of $\oc(z)$. Then $f^*\approx \frac12N \o \ob(\xi^*,0) /\E^*$ and $\tilde f\approx \frac12 \o  N/\sqrt{\E\E^*}$. 
The plateau establishes on a faster time-scale than the total number of particles, thus one may use non-stationary $N(t)$ in this formula as long as considering processes slower than $\tau_\ind{ql}$ but faster than $\tuf$. On time scales slower than $\tuf$  the losses become important. 

The regimes of weak and strong diffusion are both characterized by similar energy distributions of lost electrons: its energy is localized around maximal heating energy $\E^*$. The spread may be defined either by not-included kinetic processes, such as drifts across a magnetic field or Coulomb collisions, or by the dispersion  of the initial conditions for accelerated electrons.

 The dispersion of the initial conditions may be accounted in a simple way by matching the distribution function of fast electrons to the distribution function $\tilde{f}_\ind{s}$ of seed  electrons of bulk (cold, background) plasma. This condition, formulated at low energies, defines the spread over $\K$ in the whole relativistic domain. From a mathematical point, we have to substitute $\delta(\K)$  in \eref{eqQ4} to a function 
$\Delta(\K)$ with finite width, $\Delta(\K)\propto\int_{\xi_{\min}}^{\infty}\tilde{f}_\ind{s}(\xi,\K)\,\rmd\xi$ where the proportionality constant is defined by $\int \Delta\,\rmd\K=1$, and integration is done over $\xi$-domain consistent with \eref{eqK} and \eref{Econ}. We also restrict the integration limits only to particles with cold resonance $\o=s\oc(z)$ met somewhere along the bounce orbit, other particles can not be heated. These considerations result in $\xi_{\min}=\K$ for $\K>0$ and $\xi_{\min}=|\K|+2\sqrt{mc^2|\K|}$ for $\K<0$.  Let us assume for definiteness a Maxwellian distribution of seed electrons, $\tilde{f}_\ind{s}\propto\exp(- \E/T_\ind{e})$. 
Then 
\begin{equation}\label{eq34}\Delta(\K)=\frac2{T_\ind{e}}\cases{\exp\left(-\frac{2\K}{T_\ind{e}}\right)\!\!\!& if $\K>0$ \\ \exp\left(-\frac{2\sqrt{mc^2|\K|}}{ T_\ind{e}}\right) \!\!\!& if $\K<0$}\end{equation}  
Here we neglect terms of the order of $T_\ind{e}/mc^2\ll1$ in the pre-exponential factor.

It is important to note that regimes similar to strong diffusion are also possible for any ratio between  ${\tau_\ind{ql}}$ and ${\tuf}$ when the diffusion operator has zeros inside the confinement region, $D_\ind{ql}(\xi^{**})=0$ with $\xi^{**}<\xi^{*}$, providing zero particle flux though the point $\xi^{**}$. In particular, this is typical for resonant heating of rarefied plasma considered in \sref{res} and so-called regime of super-adiabatic limit \cite{be19}. In all these cases the energy $\E^{**}$ is not directly linked to the energy distribution of precipitating electrons.

\section{Two frequency heating}\label{tfh}

\begin{figure*}[tb]
\centering \includegraphics[width=0.33 \textwidth]{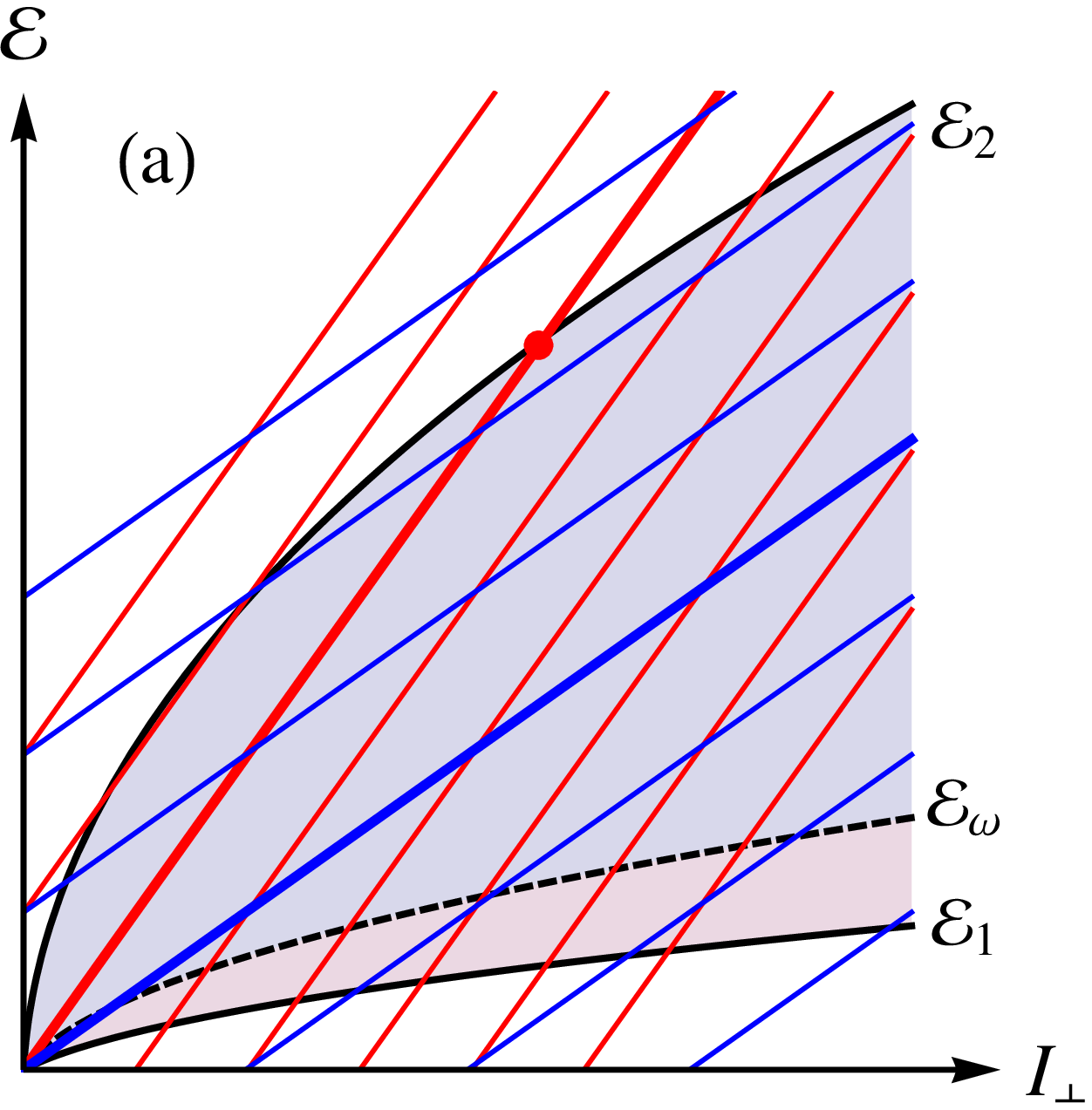}\hspace{1cm}\includegraphics[width=0.33 \textwidth]{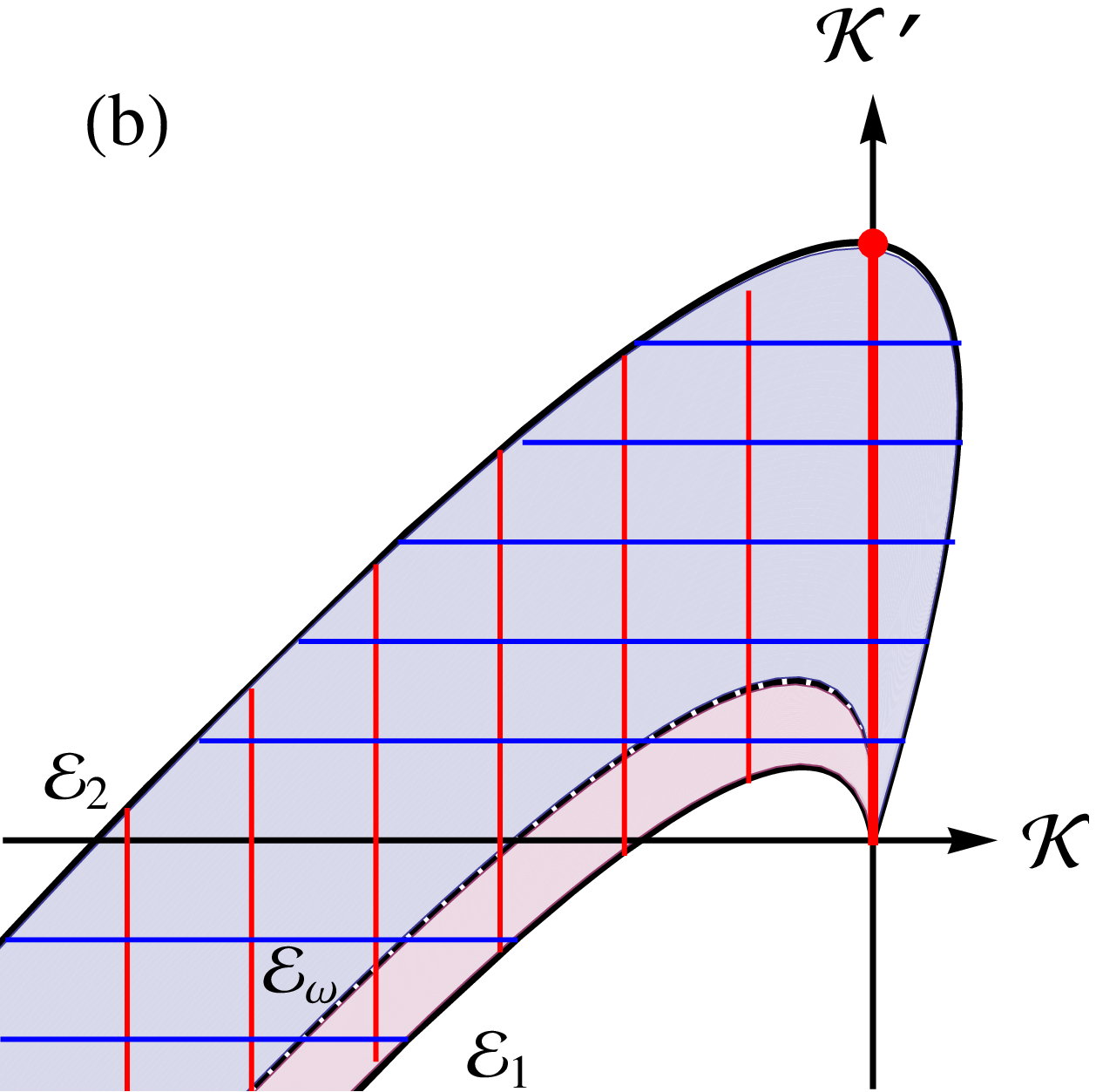}
\caption{Quasilinear diffusion curves $\K=\mathrm{const}$ (red lines, correspond to a higher frequency) and $\K'=\mathrm{const}$ (blue lines).} \label{fig-3}
\end{figure*}

Bichromatic electron-cyclotron heating may be described in a similar way --- let us append the quasilinear equation \eref{eqDQ1} with the source and loss terms and consider the resulting stationary solution. As already mentioned, two one-dimensional diffusion operators acting together result in a two-dimensional diffusion. Figure \ref{fig-3}(a) gives a naive interpretation of this process: diffusion becomes global because  one-dimensional diffusion lines  $\K,\K'=\mathrm{const}$ form a two-dimensional net in the phase-space $(\E,\I)$. 

The form \eref{eqDQ1} is convenient for such physical interpretation but not strictly defined in a mathematical sense since here we formally use a set of four variables $\xi,\xi',\K,\K'$ instead of two. To avoid uncertainties, let us choose $\K$ and $\K'$ as new independent variables instead of $\E$ and $\I$ :
\begin{eqnarray*}
\K=(\E-\o\I)/2,\qquad &\E=(\o'\K-\o\K')/\alpha,\\
\K'=(\E-\o'\I)/2,\qquad &\I=(\K-\K')/\alpha,
\end{eqnarray*}
and $\alpha=\frac12(\o'-\o)\ne 0$. With new variables both quasilinear operators may be diagonalized simultaneously. Indeed, we find that  
\begin{eqnarray*} {\o\,}\pdervi{}{\E}+\pdervi{}{\I}=-\alpha\,\pdervi{}{\K'},\\{\o'}\pdervi{}{\E}+\pdervi{}{\I}=\alpha\,\pdervi{}{\K}.\end{eqnarray*}
Then \eref{eqDQ1} takes a fully symmetric form
\begin{eqnarray*}\left[\pderv{\tilde f}{t}\right]_\mathrm{ql}=\\ \quad=\pderv{}{\K'}\left(\frac{\alpha^2D_\ind{ql}}{\ob\o^2}\pderv{(\ob\tilde f)}{\K'}\right)+\pderv{}{\K}\left(\frac{\alpha^2D'_\ind{ql}}{\ob\o'^2}\pderv{(\ob\tilde f)}{\K}\right),\end{eqnarray*}
and corresponding stationary equation is
\begin{equation}\label{eqDQ2}
\left[\pderv{\tilde f}{t}\right]_\mathrm{ql}+S-L=0.\end{equation}
These are conventional two-dimensional diffusion equations (with no mixed partial derivatives) for the distribution function $\tilde f(\K,\K')\equiv f(\E,\I)$  with the elementary volume 
$\rmd\Gamma=\rmd\K\rmd\K'/|\alpha|$. The above equations allow a nice interpretation---adding an auxiliary heating frequency results in a diffusion of diffusion lines for the primary frequency.

Analytical solution of \eref{eqDQ2} is complicated due to rather complex boundary of domain of definition of $\tilde f(\K,\K')$, see figure \ref{fig-3}(b).  Nevertheless, it is clear that one should expect much broader and smoother $\tilde f$ than singular one-dimensional distributions typical of monochromatic heating. Corresponding energy spectrum of lost electrons is also broader. The domain of definition of our solution, $\E_1(\I)\le\E\le\E_2(\I)$, is open to infinity---therefore, a stationary quasilinear plateau is not possible for general diffusion coefficients $D_\ind{ql}$ and $D'_\ind{ql}$ (the plateau may appear, however, when both diffusion coefficients are zero along a line that cuts some closed domain).  
\new{$\qquad\qquad$ 5 $>$} 
In principle, particles may be accelerated up to much higher energies than the limit $\E^*$ defined for the monochromatic heating. 

\begin{figure}[tb]
\centering \includegraphics[width=0.4 \textwidth]{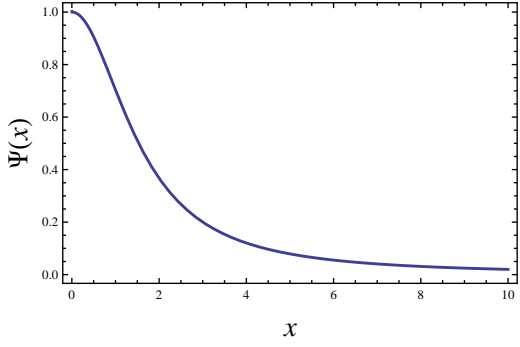}
\caption{Effect of electron distribution broadening due to presence of the second heating frequency: the total number of fast electrons normalized over the value for the monochromatic heating as a function of $x\sim({\E^*}/{\Delta\K})({{D'_\ind{ql}}/{D_\ind{ql}}})^{1/2}$. } \label{fig-g}
\end{figure}

Although complete analysis of all possible solutions of \eref{eqDQ2} is still an open issue, we consider an important particular example. Let us assume that:
\begin{itemize}
	\item the second frequency $\o'$ acts as a perturbation to the heating at the main frequency $\o$;
	\item the quasilinear diffusion is in the weak regime, i.e., the loss-cone is empty.
\end{itemize}
In zero approximation we have nearly one-dimensional distribution along $\K'$ indicated by the thick red stripe in figure \ref{fig-3}(b). This distrbution may be found in a similar way as \eref{eqQ4a}. The width along $\K$ is defined by the source of seed electrons, $\Delta\K\approx T_\ind{e}/2$, see \eref{eq34}. For the empty loss-cone, we may assume $\tilde f\approx0$ at the boundary $\E_2$ and, therefore, at each diffusion line not resting on the source, i.e. \emph{everywhere outside the thick red stripe}. Heating at the second frequency  drives the particles across that stripe towards the empty region. Naturally, it may be interpreted as additional losses. For example, one may define $\tilde n(\K') =\int \tilde f\;\rmd\K$; then for constant diffusion coefficients and bounce frequency   \eref{eqDQ2} leads to
\begin{equation*}
\frac{\alpha^2D_\ind{ql}}{\o^2}\pderv{^2\tilde n}{\K'^2}-\frac{\tilde n}{\tau'}+S_0\delta(\K')=0,\; \frac{1}{\tau'}=\frac{\alpha^2D'_\ind{ql}}{\o'^2\Delta\K^2}\end{equation*}
with an obvious stationary solution,
\begin{equation}\label{eqDQ44}
\tilde n(\K')=\frac{\o^2S_0}{\alpha^2D_\ind{ql}}\;\frac{\sinh\:[\lambda'(\K'^*-\K')]}{\lambda'\cosh(\lambda'\K'^*)},\end{equation}
where $K'^*=-\alpha \E^*/\o$  corresponds to values at the loss-cone boundary, the condition of $\tilde n(K'^*)=0$ is implied, and
\begin{equation*}
\lambda'=\frac{\o}{|\alpha|\sqrt{D_\ind{ql}\tau'}}=\frac{\o}{\o'}\sqrt{{D'_\ind{ql}}/{D_\ind{ql}}}\frac{1}{\Delta\K}\end{equation*}
characterizes the relative strength of the second frequency heating. With $\lambda'\to0$,  solution \eref{eqDQ44} defines a distribution function due to the monochromatic heating that is fully equivalent to \eref{eqQ4a}. One can find that adding the second frequency always reduces the number of accelerated particles. For instance, the total number of trapped particles is
\begin{equation*}N=\int_0^{K'^*}\tilde n(\K')\,\rmd\K'=\frac{S_0}{2 D_\ind{ql}}(\E^*)^2\,\Psi(\lambda'\K'^*),\end{equation*}
where $\Psi(x)=2(1-\mathrm{sech}\: x)/x^2$ is a monotonic function decreasing  as $2/x^2$ for large arguments; $\Psi(0)=1$ corresponds to the monochromatic heating. This function is shown in \fref{fig-g}. Note that 
\begin{equation*}x=\lambda'\K'^*\approx\frac{\o-\o'}{\o'}\:\frac{\E^*}{T_\ind{e}}\:\sqrt{{D'_\ind{ql}}/{D_\ind{ql}}}\:\end{equation*} 
may indeed be large since ${\E^*}/{T_\ind{e}}$ may  take very large values, thus the effect of smoothing of the one-dimensional distribution function by the secondary diffusion is also expected to be essential. 

Presented simple analysis allows us to estimate a sufficient condition when adding of the second frequency in the heating spectrum is not essential: $x\ll1$. The answer for the necessary condition, i.e.\ how large should be $x$ to ensure the stabilizing effect of the two-frequency heating, depends on particular mechanism of kinetic instability and its influence on the ambipolar potential confining high charge state ions. This is ongoing study that will be reported elsewhere.

\section{Cyclotron resonance condition}\label{res}

In a mirror magnetic configuration, the cyclotron resonance condition with a quasi-monocromatic wave is provided by a spatial variation of the gyrofrequency along the particle drift trajectory. 
Acceleration of electrons would change the resonant value of the qyrofrequency and, correspondingly, shift the resonance spatial position or even bring particles out of resonance. In this section we consider the consistency  between the  cyclotron resonance condition and the quasilinear diffusion flux for  ECR ion sources. 

The cyclotron resonance at the fundamental harmonic is defined by
\begin{equation*}\left\{\eqalign{ \o- k_z v_z- \gamma\,\oc=0, \cr \E+mc^2= mc^2\gamma =\sqrt{m^2c^4+c^2\gamma^2v_z^2 +2mc^2 \oc\vphantom{^1}\,\I}.}\right.
\end{equation*}
Assuming that the cold cyclotron resonance condition is met somewhere along the field line, see \eref{wcon}, we can exclude varying $\oc$  and formulate a quadratic equation for the resonant velocity,
\begin{equation*}\beta_z^2-\frac{2n_z\o\I}{mc^2\gamma}\beta_z+\frac{2\o\I}{mc^2\gamma}+\frac1{\gamma^2}-1=0,\end{equation*} 
where $\beta_z=v_z/c$ and $n_z=c k_z/\o$ is the longitudinal refractive index at the cyclotron resonance position.
Real solutions for $\beta_z$ exist when
\begin{equation*} \left(\frac{\o\I}{mc^2}-\gamma\right)^2\ge 1+(1-n_z^2)\left(\frac{\o\I}{mc^2}\right)^2.\end{equation*}
Next we combine this with the relation for the quasilinear diffusion line
\begin{equation*} \K=mc^2(\gamma-1)-\o\I=\mathrm{const}\end{equation*}
and obtain
\begin{equation} \label{eq33}
\left(\frac{\K}{mc^2}+1\right)^2\ge 1+(1-n_z^2)\left(\frac{\o\I}{mc^2}\right)^2.
\end{equation}

Let us consider first  $n_z<1$, usually referred as the case of \emph{rarefied plasma}. Inequality \eref{eq33} then defines the upper limit energy of the resonant particle,
\begin{equation*}\E\le\E^{**}={\K}+\sqrt{\frac{(\K+mc^2)^2-m^2c^4}{1-n_z^2}}.\end{equation*}
Opposite to our previous assumption, a particle with zero energy can not gain energy since $\E^{**}=0$ for $\K=0$. So to estimate maximal energy in this case,  one should assume some dispersion $T_\ind{e}$ over initial energies, which results in a spread of $T_\ind{e}/2$ over $\K$, see \eref{eq34}.  If $T_\ind{e}\ll mc^2$, \begin{equation*}\E^{**}\approx\sqrt{mc^2T_\ind{e}/(1-n_z^2)}\ll \E^*\sim mc^2,\end{equation*} i.e.\ cyclotron interaction stops before the resonant particle is accelerated to relativistic energies that are needed to push the particle into the loss-cone. In this case, a quasilinear plateau may be formed inside the confinement region of a phase space.

The opposite case with $n_z>1$, the case of \emph{dense plasma}, is much more typical of modern ERC ion sources operating at the fundamental electron-cyclotron harmonic. Indeed, in the presence of dense enough cold background plasma the dispersion relation for the right-hand polarized waves launched presumably along the magnetic field may be approximated by a whistler-like dependence,  \cite{Sazhin} 
\begin{equation*} n_z^2=1-\o_\ind{p}^2/\o(\o-\oc),\end{equation*} where $\o_\ind{p}$ is an electron Langmuir frequency of the background  plasma. From this equation we see than once the cold cyclotron resonance condition at the fundamental harmonic, $\o=\oc$, is met somewhere along the drift trajectory, the refractive index may acquire infinitely large values. Correspondingly, condition \eref{eq33} may be fulfilled for any energy $\E$. In the same manner as \eref{Econ}, we find that particles with the kinetic energy equal to \begin{equation*}\E_\o=\sqrt{m^2c^4+2mc^2 \o\vphantom{^1}\,\I}-mc^2,\end{equation*} have turning points exactly at $\o=\oc$, and particles with $\E>\E_\o$ meet the cold resonance before being reflected by the magnetic mirror. Hence, all particles satisfying 
\begin{equation*}\label{Econ1_}\E_\o\le\E\le\E_2
\end{equation*} 
can be accelerated in a dense plasma without losing the cyclotron resonance condition, compare to \eref{Econ}. The boundary $\E=\E_\o$ is indicated in figures \ref{fig-2} and \ref{fig-3} by different filling. Note that $\E_\o<\o\I$, i.e. if a particle accelerates at the fundamental harmonic from zero energy, it is always in resonance, see figure \ref{fig-2};  at higher harmonics this is not true.
Above we consider the sufficient conditions of the cyclotron resonance in a dense plasma. More accurate analysis reveals that they  are rather close to necessary conditions in the parameter range relevant to modern experiments. 

A characteristic density that separates the rarefied and dense cases may be estimated from the following fact. Components of the dielectric tensor for  waves in thermal plasma with Langmuir frequency $\o_\ind{p}$ and temperature $T_\ind{e}$ can not deviate essentially from unity if condition 
$\o_\ind{p}^2/\oc^2\ll \sqrt{T_\ind{e}/mc^2}$ is fulfilled \cite{Sazhin,be22}. In all other cases excitation of slow waves with big refractive index is always possible near the cold cyclotron resonance \cite{egor}.  
\new{$\qquad\qquad$ 4 $>$} 
Thus, $\o_\ind{p}^2/\oc^2\gtrsim \sqrt{T_\ind{e}/mc^2}$ may be considered, at first approximation, as a boundary between the rarefied and dense plasma cases. With the bulk electron temperature of few tens or hundreds eVs, the dense plasma regime is likely realized in most modern experiments. Note that the ``dense plasma'' may be far from the cutoff density, i.e.  $\sqrt{T_\ind{e}/mc^2}<\o_\ind{p}^2/\oc^2\ll 1$; this means that plasma is transparent for all electromagnetic waves except those with an electric field rotating in phase with the electrons.

In modern high-performance ECR ion sources the dense plasma condition, \begin{equation}\label{eqd}\sqrt{T_\ind{e}/mc^2}\ll\o_\ind{p}^2/\oc^2\sim1,\end{equation} is usually met with large extend. However, there is an important exception for which rarefied plasma limit works despite condition \eref{eqd} still holds. This is the case when the wave frequency is below the cyclotron frequency in the entire trap volume. In this case the cold plasma resonance is never met, thus making wave slowing down impossible. In experiments, this non-resonant wave is a second wave acting together with the main (resonant) heating wave; it may be either external wave in two-frequency heating experiments \cite{izo1,izo2,izo3,olli_2015}, or a cavity mode excited internally by a strongly nonequlibrium electron population \cite{izot-fe}.

\section{Summary}

In the present communication we report the first steps made towards understanding the physics of two-frequency heating in modern ECR ion sources. Within the frame of the quasilinear model of plasma-wave interaction,  we assume that the second heating wave may essentially expand the domain of random Brownian motion of resonant electons in a phase space. Thus, we contrast the distribution functions of fast electrons driven by monochromatic and bicromatic ECR heating and show , on a rather qualitative level, that in the latter case one should expect much mode smooth and less populated electron distributions. A kinetic stability analysis of the resulted distributions is an obvious next step to be reported in a forthcoming publication.

\section*{Acknowledgments}
The work was supported by the Russian Science Foundation, project No~19-12-00377.

\section*{References}

\end{document}